\documentclass[twocolumn, showpacs, english, preprintnumbers, amsmath, amssymb, superscriptaddress, aps]{revtex4-1}
\usepackage{latexsym}
\usepackage{epsfig,psfrag}
\usepackage{dcolumn}
\usepackage{bm}
\usepackage{graphicx}
\usepackage{color}
\usepackage{esint}
\usepackage{babel}
\usepackage{amsfonts}
\usepackage{enumerate}
\usepackage{mathbbol}

\begin{document}
\title{Interaction-induced conductance from zero modes in a clean magnetic graphene waveguide} 

\author{Laura Cohnitz}
\affiliation{Institut f\"ur Theoretische Physik, 
Heinrich-Heine-Universit\"at, D-40225 D\"usseldorf, Germany}

\author{Wolfgang H\"ausler}
\affiliation{Institut f\"ur Physik, 
Universit\"at Augsburg, D-86135 Augsburg, Germany}
\affiliation{I. Institut f\"ur Theoretische Physik, 
Universit\"at Hamburg, D-20355 Hamburg, Germany}

\author{Alex Zazunov}
\affiliation{Institut f\"ur Theoretische Physik, 
Heinrich-Heine-Universit\"at, D-40225 D\"usseldorf, Germany}

\author{Reinhold Egger}
\affiliation{Institut f\"ur Theoretische Physik, 
Heinrich-Heine-Universit\"at, D-40225 D\"usseldorf, Germany}

\date{\today}

\begin{abstract}
We consider a waveguide formed in a clean graphene monolayer by a
spatially inhomogeneous magnetic field.  
The single-particle dispersion relation for this waveguide exhibits
a zero-energy Landau-like flat band, while finite-energy bands have 
dispersion and correspond, in particular, to snake orbits. 
For zero-mode states, all matrix elements of the current operator  
vanish, and a finite conductance 
can only be caused by virtual transitions to finite-energy bands. 
We show that Coulomb interactions generate such processes.
In stark contrast to finite-energy bands, the conductance is not 
quantized and shows a characteristic dependence on the zero-mode filling.
Transport experiments thereby offer a novel and highly sensitive probe of 
electron-electron interactions in clean graphene samples.
We argue that this interaction-driven zero-mode conductor 
may also appear in other physical settings and is not captured 
by the conventional Tomonaga-Luttinger liquid description.
\end{abstract}


\maketitle

\section{Introduction}\label{sec1}

Electronic phases exhibiting flat bands have attracted considerable
attention over the past few decades \cite{flatrev1,flatrev2,flatrev3}. 
For instance, flat bands can arise from interference effects on a
geometrically frustrated lattice.  On the noninteracting level, due to the lack of dispersion, one expects insulating behavior when the Fermi 
level resides inside the flat band, such that the conductance vanishes identically at zero temperature.  For lattice models hosting almost flat bands, it is well known that interactions can cause dramatic effects such as topologically nontrivial fractional Chern insulator 
phases \cite{wen,dassarma,chamon}. Even topologically trivial phases without any dispersion can show remarkable behavior.  For instance, in the case of long-range unscreened interactions, the conductance of the so-called ${\cal T}_3$ lattice can be finite and exhibits a highly nontrivial dependence on the filling factor \cite{hausler}. 
Somewhat related conclusions have been obtained for interacting fermions with short-range interactions on lattices with geometrically
frustrated unit cells, where at certain filling factors
the noninteracting theory predicts insulating behavior but repulsive 
interactions cause the existence of delocalized two-particle states 
\cite{doucot1,doucot2,doucot3,movilla,lopes,lopes2}.
Such effects have been studied in detail for diamond chains, 
where flat band formation emerges due to Aharonov-Bohm caging. 
However, two-particle delocalization does not necessarily imply that the 
many-particle electron system will have finite conductance at finite density  \cite{doucot1,doucot2}.  Similar issues have also been discussed
for interacting bosons \cite{takayoshi} and for cold-atom systems \cite{scarola}.

In this work, we show that a finite conductance is generated
by Coulomb interactions in another flat-band system, referred to as
``magnetic graphene waveguide'' (MGW) in what follows.  
Our analysis for the MGW reveals that interactions can turn a 
noninteracting insulator into a conductor, even though 
electron-electron interactions usually suppress the 
conductance \cite{aa,zala,kupfer}.
This effectively one-dimensional (1D) zero-mode conductor falls 
outside the conventional Tomonaga-Luttinger liquid (TLL) description of 
interacting 1D conductors \cite{gogolin-book}.  We expect that such a state 
also appears in other physical 
settings and provide an in-depth description of its 
properties in a MGW.

\begin{figure}
\centering
\includegraphics[width=5.5cm]{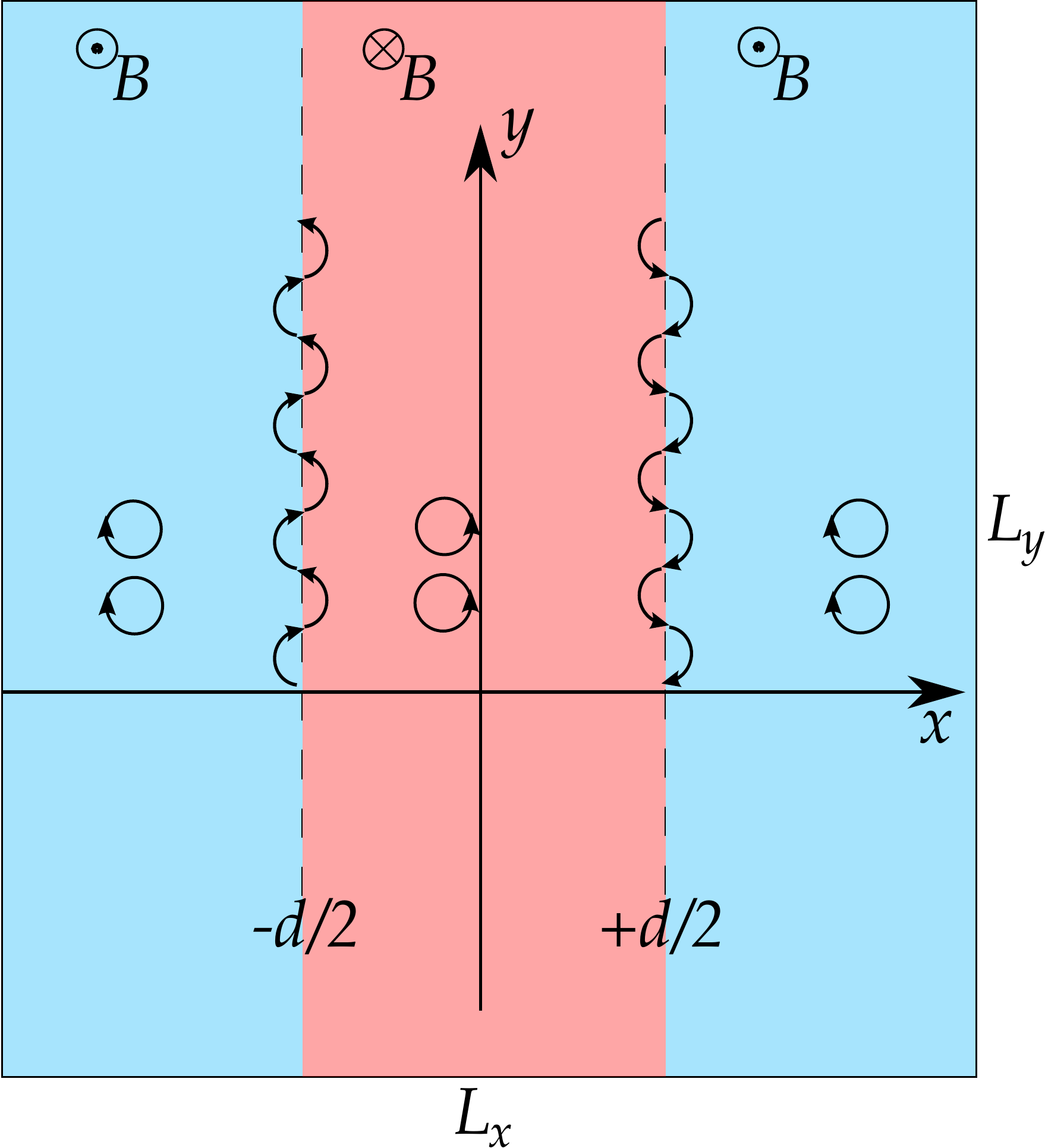}
\caption{\label{fig1}
 Schematic sketch of the MGW setup viewed from above.  
A magnetic field $+B\hat{\bm e}_z$ is applied everywhere in the graphene plane 
except for the waveguide region $|x|<d/2$, where the field is reversed. Near the 
field switching lines, quantum modes corresponding to classical snake orbits
are realized, while only Landau-quantized cyclotron orbits are present far 
away from these lines.  Several classical orbits corresponding to $n=1$ are 
schematically illustrated. The system dimensions are $L_x$ and $L_y$, respectively. 
}
\end{figure}

Our MGW setup is illustrated in Fig.~\ref{fig1}, where a clean graphene monolayer is exposed to a static inhomogeneous magnetic field.  Very long mean free paths have already been realized in graphene, e.g., by using boron nitride as substrate \cite{boron}.  Low-energy quasiparticles close to the neutrality point then correspond to massless Dirac fermions in two spatial dimensions (2D) \cite{rmp1,rmp2}. 
The magnetic field is taken spatially inhomogeneous along, say, 
the $x$-direction, ${\bm B} = B(x) \hat {\bm e}_z$, where we
focus on orbital fields such that only the perpendicular ($z$-)component 
matters.  To be specific, we study the field profile, see Fig.~\ref{fig1}, 
\begin{equation}\label{magprof}
B(x)= B\ {\rm sgn}(|x|-d/2)
\end{equation}
describing a MGW of transverse width $d$ where the 
magnetic field is reversed ($B\to -B$) in the waveguide region defined by $|x|<d/2$.
Such inhomogeneous magnetic field profiles allow one to guide 
Dirac fermions \cite{ademarti,ademarti2,peeters,magnscatt,ghosh,kuru}, 
and single-particle spectra of the resulting MGW
have been analyzed in detail \cite{lambert,tarun,hausler2,bliokh,prada}.
For $d\to 0$, Eq.~\eqref{magprof} reduces to the familiar homogeneous field
case and one recovers the well-known dispersionless relativistic Landau levels,
including a zero mode \cite{gusynin,gusynin2,goerbig,miransky}. 
Importantly, such a zero-energy 
band (with band index $n=0$) is also present for finite $d$, 
forming the flat band of interest below, while all other ($n\ne 0$) bands
acquire dispersion.  
In particular, near $x=\pm d/2$, pairs of counterpropagating 
``snake states'' develop \cite{lambert,tarun,sim,artur,shakouri}, which are either of electron ($n>0$) or hole $(n<0)$ type. Classical snake orbits forming near zero-field lines are illustrated schematically in Fig.~\ref{fig1}. We note that snake states have already been observed experimentally in graphene \cite{marcus,ozyilmaz,christian}, including studies of the ballistic (disorder-free) 
limit \cite{christian}.

When the $n=0$ level is partially filled, with all negative-energy
states occupied, the noninteracting conductance of the MGW vanishes 
identically since all $n=0$ current matrix elements are zero. 
We here demonstrate that, nonetheless, a finite conductance emerges due to
 inter-band Coulomb interactions (corresponding to Landau level 
mixing for $d=0$ \cite{macdon}).  The transport features related
to the zero-energy band differ qualitatively from those of  
finite-energy bands, and thus should be easy to identify experimentally.

For the guidance of the focused reader, we now summarize the content of the 
following sections before starting with the detailed description of our 
results. We describe the solution 
of the single-particle problem in Sec.~\ref{sec2a}, where in 
Landau gauge the problem is homogeneous along the $y$-direction and 
eigenstates for given band index $n$ are classified by the momentum $k_y=k$.
Coulomb matrix elements are discussed in Sec.~\ref{sec2b}, followed by 
a study of the theory projected to the zero-mode sector in
 Sec.~\ref{sec3}.  Employing the 
Hartree-Fock (HF) approximation, we find that the spatial 
inhomogeneity of the magnetic field generically leads to dispersion of
the HF single-particle energies $\varepsilon_k$.  
The HF ground state represents a filled Fermi sea, 
where all $n=0$ states with $|k|<k_F(\nu)$ are occupied, 
with the zero-mode filling factor $\nu$.
We determine the interaction-induced Fermi momentum $k_F(\nu)$, 
and show that the conductance still vanishes within the  
zero-mode description.  In Sec.~\ref{sec4}, the zero-temperature linear 
conductance, $G$, will be computed from the Kubo formula,  taking 
into account inter-band interactions  through a systematic 
perturbative expansion up to second order.  For the zero-energy bands,
we find completely different transport properties when compared to
finite-energy bands.  In the latter case, which has been studied in Ref.~\cite{hausler2}, $G$ is independent of the band filling and quantized in units of the 
conductance quantum $G_0=e^2/h$. Moreover, finite-temperature corrections to 
this quantized value follow the predictions of TLL theory. 
In marked contrast to these finite-energy bands, the conductance found 
for the zero-energy case strongly depends on the filling and is 
not described by TLL theory.  Finally, in Sec.~\ref{sec5}, we shall put our 
results into a general context. 
Some calculational details can be found in the Appendix. 
Throughout this paper, we focus on the most 
interesting zero-temperature limit and often employ units with $\hbar=1$.

\section{Magnetic graphene waveguide model}\label{sec2}

\subsection{Single-particle description}\label{sec2a}

We begin by discussing the single-particle description of the MGW, 
see also Refs.~\cite{lambert,tarun}.  
The electronic low-energy physics in a weakly doped graphene 
monolayer is well described by massless 2D Dirac fermions \cite{rmp1}.  
Including the vector potential ${\bm A}$ encoding a static 
inhomogeneous orbital magnetic field, where only the field 
component perpendicular to the layer matters, 
${\bm B}=B(x)\hat{\bm e}_z$,  the single-particle Hamiltonian is given by
\begin{equation} \label{h0}
H_0  =v_F {\bm \sigma}\cdot \left( -i{\bm\nabla} + 
\frac{e}{c} {\bm A} \right) ,
\end{equation}
where $v_F\approx 10^6$~m$/$s is the Fermi velocity, ${\bm\sigma}=(
\sigma_x,\sigma_y)$ contains $2\times 2$ Pauli matrices,
and ${\bm\nabla} =(\partial_x,\partial_y)$.
The Pauli matrices act in the sublattice space corresponding to the two 
carbon atoms in the basis of graphene's honeycomb lattice. 
The lengthscale on which $B(x)$ changes is assumed large 
against the lattice spacing ($2.46$~\AA), such that valley ($K$ point) mixing is irrelevant.  The inclusion of spin and/or valley degrees of freedom is left to future work, although this step is not expected to 
significantly affect any of our conclusions.  
As a consequence, we neglect the spin and valley degrees
of freedom in Eq.~\eqref{h0} and focus on a single Dirac cone.

We consider the magnetic field profile in Eq.~(\ref{magprof}), describing a 
MGW of width $d$, with constant field $B(x)=B$ everywhere except 
in the strip $|x|<d/2$ where $B\to -B$, see Fig.~\ref{fig1}.  
Using the Landau gauge, the vector potential is given by
\begin{equation}\label{vecpot}
{\bm A} = A(x) \hat{\bm e}_y,\quad A(x) = B\times \left\{ \begin{array}{cc}
x+d, & x<-d/2,\\ -x, & |x|<d/2,\\ x-d,& x>d/2.\end{array}\right.
\end{equation}
The specific form in Eq.~(\ref{magprof}) has been adopted in order to keep
 the discussion focused and to allow for analytical progress. 
However, other field profiles creating such a 
MGW are expected to yield similar results \cite{tarun}, and indeed it is not
necessary (nor even desirable) to have atomically sharp changes in 
the magnetic field profile.

Below we often measure lengths (energies) in units of the magnetic length $l_B$ (magnetic energy $E_B$), 
\begin{equation}\label{units}
l_B=\sqrt{\hbar c/eB}, \quad E_B=\hbar v_F/l_B.
\end{equation} 
We mention in passing that instead of a true magnetic field, 
one could also employ strain-induced pseudo-magnetic fields \cite{strain,haas},
and with minor modifications, similar physics can also be realized by employing the surface states of 3D topological insulators \cite{hasan,rosch}.  

Periodic boundary conditions along the $y$-direction quantize
the conserved momentum $k=k_y$ along the MGW.
Spinor eigenstates, $\Psi_{n,k}(x,y)$, solving the Dirac equation,
\begin{equation}\label{dirac}
H_0 \Psi_{n,k} = E_{n,k} \Psi_{n,k},
\end{equation}
with eigenenergy $E_{n,k}$ can thus be classified by $k$
and the integer band index $n$.
For a homogeneous field ($d=0$), Eqs.~\eqref{h0} and \eqref{dirac} lead to 
the well-known $k$-independent relativistic Landau 
levels \cite{rmp1,gusynin,gusynin2,goerbig,miransky}, 
$E_{n,k}^{(d=0)} = {\rm sgn}(n) \sqrt{2|n|},$
and $n$ can be identified with the Landau level index. 
When $d$ is finite, single-particle states in general exhibit 
dispersion, but bands with different band index $n$  
remain always separated by a finite gap \cite{lambert,tarun}. 
Moreover, even in this inhomogeneous case, a flat zero-energy 
band is present, $E_{n=0,k}=0$.   In this work, we study whether
this zero-mode band can carry electric current when Coulomb interaction effects
are taken into account.

Owing to momentum conservation along the $y$-direction, we have
\begin{equation}\label{free-solution}
\Psi_{n,k}(x,y) = \frac{e^{iky}}{\sqrt{L_y}} \psi_{n,k}(x) ,
\quad \psi_{n,k}(x)= \left(\begin{array}{c} \phi_{n,k}(x) \\ i \chi_{n,k}(x)
\end{array} \right),
\end{equation}
with the normalization condition $\int dx (\phi^2_{n,k}+\chi^2_{n,k})=1$.
As shown in App.~\ref{appa}, both $\phi_{n,k}$ and $\chi_{n,k}$ 
can be chosen real-valued, and the single-particle problem
is solved by matching the respective spinor solutions in the three regions 
in Fig.~\ref{fig1} at $x=\pm d/2$. The solutions have the symmetry properties
\begin{eqnarray}\label{symrel2}
\psi_{n,k}(x) &=& \sigma_z\psi_{-n,k}(x) \quad (n\ne 0),\\
\psi_{n,k}(x) &=& (-1)^{n+1}\psi^\ast_{n,-k}(-x) ,\nonumber
\end{eqnarray}
 where ``$\ast$'' indicates complex conjugation of both spinor entries.
The first relation says that a particle-hole transformation 
connects states with $n$ and $-n$ (for $n\ne 0$), and thus
the single-particle spectrum is mirror-symmetric
around zero energy, $E_{-n,k}=-E_{n,k}$.  
The second relation follows from inversion symmetry
together with the node rule in 1D.  
Moreover, the matrix elements of the current $I$ (evaluated at $y=0$) 
along the waveguide ($y$-)direction are given by
\begin{equation}\label{currmatrel}
I_{n,k;n'k'} = v_F \int dx\ \psi_{n,k}^\dagger(x) \sigma_y  
\psi^{}_{n',k'}(x),
\end{equation}
where the dagger denotes transposition and complex
conjugation of spinor wave functions \cite{footcurr}.
As a consequence of Eq.~\eqref{symrel2}, they
satisfy the symmetry relations
\begin{eqnarray}\label{cursymrel}
I_{n,k;-n,k'} &=& -I_{n,k';-n,k},\\ \nonumber
I_{0,k;n,k'} & =& (-1)^{n+1} I_{0,-k;n,-k'},
\end{eqnarray}
with arbitrary integer $n$.   

\begin{figure}
\centering
\includegraphics[trim = 0 10 80 80,clip,width=8cm]{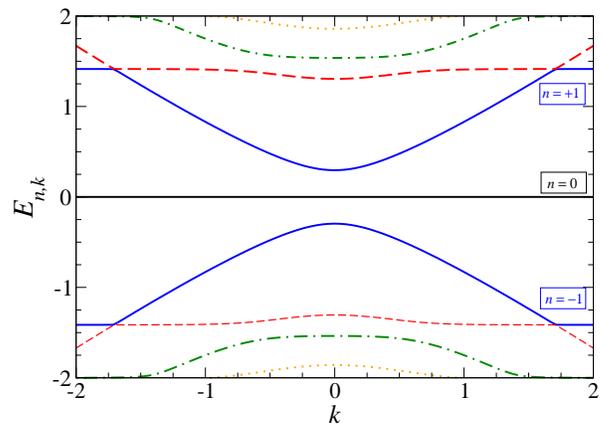}
\caption{\label{fig2}  Dispersion relation of the MGW for $d=2l_B$.
Only the marked three bands ($n=-1,0,1$) will be kept later on 
in Sec.~\ref{sec4c}, assuming that the zero mode is partially filled.  
We note that the $n=1$ and $n=2$ bands are separated by an avoided
crossing not apparent on the shown scale. 
Energy (momentum) is given in units of $E_B$ ($l_B^{-1}$), see
Eq.~\eqref{units}. }
\end{figure}

We next describe the spectrum and the eigenstates for the MGW, where 
the dispersion relation is shown in Fig.~\ref{fig2} for $d=2l_B$.
As described in App.~\ref{appa}, the spectrum and 
the eigenstates follow by numerical solution of an eigenvalue problem. 
However, analytical progress is possible for the zero modes.
Indeed, as dictated by index theorems \cite{goerbig,miransky}, 
there must be zero-energy states, $E_{n=0,k}=0$, 
also for finite $d$, cf.~Fig.~\ref{fig2}. 
Using units with $l_B=1$, their analytical form is given by  
\begin{eqnarray}\label{zeropsi}
\psi_{0,k}(x) &=& \left(\begin{array}{c}
0\\ i \chi_{0,k}(x) \end{array}\right),\\\nonumber
\chi_{0,k}(x) & =& \frac{ e^{d|x|-(x+k)^2/2}}{N_{0,k}} 
\times \left \{
\begin{array}{cc} 1,& |x|>d/2,\\  e^{(|x|-d/2)^2},&|x|<d/2,\end{array}\right.
\end{eqnarray}
with normalization constant $N_{0,k}$.  Due to the absence of an upper spinor 
component, all zero-mode current matrix elements vanish identically,
\begin{equation}\label{zerocur}
I_{0,k;0,k'}=0.
\end{equation}
For $d=0$, the $\chi_{0,k}(x)$ functions reduce to shifted 
harmonic oscillator ground-state 
wavefunctions describing the $n=0$ Landau level.
For $d\ne 0$, the probability density distribution, $|\psi_{0,k}(x)|^2$, 
has two local maxima near (for $|k|\alt d/l_B^2$) but never inside 
the waveguide region.  For later use, we also mention that a local 
minimum of $|\psi_{0,k}(x)|^2$ exists at $x=kl_B^2$ when $|k|<d/2l_B^2$.

Since the $n\ne 0$ (and particularly the $n=\pm 1$) snake states exhibit their
probability density maximum near the null lines of the magnetic
field \cite{tarun} at $x=\pm d/2$, cf.~Fig.~\ref{fig1},
interaction-induced transitions between $n=0$ and $n\ne 0$
states are therefore only important for $|k|\alt d/ l_B^2$.
The zero-mode conductance $G$ discussed in Sec.~\ref{sec4} is
caused by precisely such transitions.  In fact, 
as long as virtual band transitions to $n\ne 0$ states are excluded, 
$G=0$ holds on general grounds since interactions cannot generate 
an upper spinor component in Eq.~\eqref{zeropsi} from the
zero-mode sector only, cf.~Eq.~(\ref{zerocur}). 

\begin{figure}
\centering
\includegraphics[trim = 0 10 80 80,clip,width=8cm]{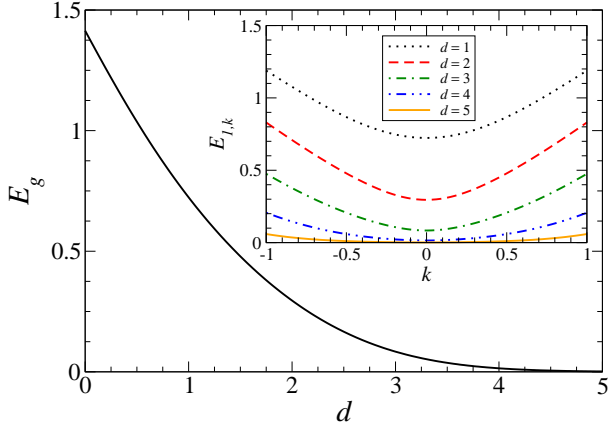}
\caption{\label{fig3}  
Main panel: Single-particle gap $E_g$ vs waveguide width $d$ in units as 
in Fig.~\ref{fig2}.  Inset: Long-wavelength part of the dispersion relation 
for $n=1$ and several values of $d$.  }
\end{figure}

Our perturbative approach in Sec.~\ref{sec4} holds as long as the single-particle gap, $E_{g} = E_{n=1,k=0}$, which separates the zero 
mode from the $n=1$ band, is large against the 
typical Coulomb energy scale (see Sec.~\ref{sec2b}). 
Figure \ref{fig3} shows $E_{g}$ as a function of the 
MGW width $d$. The observed decrease suggests to subsequently 
consider only $d\alt 2l_B$. 
Nonetheless, as shown in App.~\ref{appb}, it is also 
instructive to discuss the limit $d\to\infty$, where a rapid 
(approximately exponential) decrease of $E_{g}$ is seen in Fig.~\ref{fig3}. 

\subsection{Coulomb interactions} \label{sec2b}

Next we turn to a second-quantized description and include Coulomb interaction 
effects.  The fermion annihilation field operator at 
${\bm r}=(x,y)$ is written as
\begin{equation}\label{fieldop}
\hat\Psi({\bm r}) = \sum_{n,k} \Psi_{n,k}({\bm r})\ c_{n,k}, 
\end{equation}
with fermion operators $c_{n,k}$ subject to the standard
anticommutator algebra, $\{ c^{}_{n,k}, c_{n',k'}^\dagger\}=\delta_{nn'}
\delta_{kk'}$ and so on, and $\Psi_{n,k}$ in Eqs.~\eqref{dirac} and 
\eqref{free-solution}.  Using the units in Eq.~\eqref{units}, 
the second-quantized interaction Hamiltonian is given by 
\begin{equation}\label{hint}
\hat H_I= \frac12 \int d{\bm r}d{\bm r}' \ V({\bm r}-{\bm r}') 
\hat\Psi^\dagger( {\bm r}) \hat\Psi^\dagger( {\bm r}') \hat\Psi^{}( {\bm r}')
\hat\Psi^{}( {\bm r}), 
\end{equation}
where we consider a gate-screened Coulomb potential,
\begin{equation}\label{potential1}
V({\bm r}) = \alpha \left( \frac{1}{|{\bm r}|}
-\frac{1}{\sqrt{|{\bm r}|^2+4R^2}} \right).
\end{equation}
Here $\alpha$ denotes graphene's effective fine structure constant,
with typical values in the range $\alpha\approx 0.1$ to $2$, 
depending on the dielectric properties of the surroundings  \cite{rmp1,rmp2}.  
The second (image charge) term in Eq.~\eqref{potential1} comes from a
parallel metallic plate, e.g., due to a gate electrode, at distance $R$ 
from the graphene layer.  The Fourier transform of Eq.~\eqref{potential1} 
is given by (we remind the reader that $k=k_y$)
\begin{equation} \label{tildev}
\tilde V(k_x,k) = 2\pi \alpha 
\frac{1-e^{-2R\sqrt{k_x^2+k_{}^2}}}{\sqrt{k_x^2 +k_{}^2}}.
\end{equation}
Inserting Eq.~\eqref{fieldop} into Eq.~(\ref{hint}) and 
exploiting momentum conservation along the $y$-direction, we obtain
\begin{eqnarray}\nonumber
\hat H_I &=& \frac12 \sum_{n_1,n_2,n_3,n_4} \sum_{k,k',q}  
V_{k,k';q}^{(n_1,n_2, n_3,n_4)} \\  &\times& 
c^\dagger_{n_1,k} c^\dagger_{n_2,k'} c^{}_{n_3,k'-q} c^{}_{n_4,k+q},
\label{hint2}
\end{eqnarray}
where the Coulomb matrix elements
\begin{equation} \label{coul}
V^{( \{n_j\} )}_{k,k';q} =  \frac{1}{L_y}
\int \frac{d k_x}{2\pi} \tilde V ( k_x, q ) 
{\cal F}^{(n_1,n_4)}_{k,q}(k_x) {\cal F}^{(n_2,n_3)}_{k',-q}(-k_x)
\end{equation}
are expressed in terms of form factors,
\begin{equation}\label{bdef}
{\cal F}_{k,q}^{(n,n')}(k_x) = 
\int dx \ e^{-ik_x x} \psi^\dagger_{n,k}(x)\cdot \psi^{}_{n',k+q}(x) .
\end{equation}
We show in App.~\ref{appc} that the Coulomb matrix elements in
Eq.~\eqref{coul} are real-valued and subject to the symmetry relations
\begin{eqnarray} \label{symrel}
V^{(n_1,n_2,n_3,n_4)}_{k,k';q}&=& V^{(n_2,n_1,n_4,n_3)}_{k',k;-q}\\ \nonumber
&=& V^{(n_4,n_3,n_2,n_1)}_{k+q,k'-q;-q}\\ \nonumber
&=& (-1)^{n_1+n_2+n_3+n_4} V^{(n_1,n_2,n_3,n_4)}_{-k,-k';-q}.
\end{eqnarray}
In order to obtain numerical values for the Coulomb matrix elements, 
we first compute the form factors in Eq.~\eqref{bdef} by numerical 
integration over $x$, taking into account their symmetry properties.  
Given the form factors, the remaining $k_x$-integration in Eq.~\eqref{coul} can 
then be evaluated numerically in an efficient manner.

Notably, the zero-mode ($n=n'=0$) form factors can be evaluated analytically, 
\begin{equation} \label{bkq}
{\cal F}^{(0,0)}_{k,q}(k_x) = 
\frac{ Y\left (k+\frac{q+ik_x}{2}\right ) }
{ \sqrt{Y(k) Y(k+q)} }.
\end{equation}
Using again units with $l_B=1$, we here use the
complex-valued auxiliary function 
\begin{eqnarray} \label{akqQ}
Y(z) &=& \sum_\pm \Bigl ( e^{(z\mp d)^2} 
{\rm erfc}\left(\pm [z \mp d/2] \right) \\ \nonumber
& \pm & i e^{-z^2+d^2/2} {\rm erfc}\left(-i[z\mp d/2] \right) \Bigr),
\end{eqnarray}
with the  complementary error function 
${\rm erfc}(z)$ \cite{gradst,abramowitz}. 
For real-valued argument $z$, the function  $Y(z)$ is real and positive.
In Sec.~\ref{sec3}, we shall also refer to the
homogeneous case $d\to 0$, where Eq.~\eqref{akqQ} 
simplifies to $Y(z)\to 2e^{z^2}$. The form factors in Eq.~\eqref{bkq} then
become ${\cal F}^{(0,0)}_{k,q}(k_x) \to e^{-(k_x^2+q^2)/4} e^{i(k+q/2)k_x}$, 
resulting in the zero-mode Coulomb matrix elements
\begin{eqnarray} \label{d0zmC}
\left. V_{k,k';q}^{(0,0,0,0)}\right|_{d\to 0} &\to & 
\frac{2\alpha}{L_y}\Bigl[ K_0\left(|q(k-k'+q)|\right) \\ \nonumber
&-&K_0 \left( |q|\sqrt{4R^2+(k-k'+q)^2}\right)\Bigr]
\end{eqnarray}
with the modified Bessel function $K_0$ \cite{gradst}.

Finally, it simplifies our subsequent analysis to use antisymmetrized
Coulomb matrix elements throughout the remainder of this paper,
\begin{equation}\label{cmt1}
W_{k,k';q}^{(n_1,n_2,n_3,n_4)} = \frac12 \left(
V^{(n_1,n_2,n_3,n_4)}_{k,k';q} -V^{(n_2,n_1,n_3,n_4)}_{k',k;q+k-k'} \right),
\end{equation}
which follow from Eq.~\eqref{coul} by antisymmetrization 
under the exchange $(n_1,k)\leftrightarrow(n_2,k')$. 
This antisymmetrization simply reflects the fermionic anticommutator algebra.  
The matrix elements in Eq.~\eqref{cmt1} are also real-valued and
enjoy the same symmetry relations, see Eq.~\eqref{symrel}, as 
the $V^{(n_1,n_2,n_3,n_4)}_{k,k';q}$.

\section{Zero mode sector: Hartree-Fock theory} \label{sec3}

We now consider the case of a partially filled zero mode, 
where all negative-energy bands ($n<0$) are occupied while
all positive-energy states $(n>0)$ are unoccupied. The $n=0$ level 
has the filling factor $\nu=N/N_s$, where $N$ particles occupy 
the $n=0$ band and the degeneracy degree, $N_s$, 
is given by the total magnetic flux in units of the flux quantum, 
\begin{equation}\label{nsdef}
N_s= \frac{(L_x-2d) L_y}{2\pi l_B^2},
\end{equation}
assuming a rectangular sample, see Fig.~\ref{fig1}.  The 
momentum $k=k_y$ takes the values $k=2\pi n_y/L_y$ with $-N_s/2<n_y\le N_s/2$. 
This assumption of fully occupied (empty) bands with negative (positive) energy also holds for the interacting ground state as
long as the typical Coulomb energy scale is small compared to the 
single-particle gap $E_g$, see Fig.~\ref{fig3}.  In this section, we consider 
the zero-mode sector only and thus neglect all Coulomb interaction processes 
involving $n\ne 0$ states.  

In the zero-mode theory, there is no kinetic energy term and
the Hamiltonian equals $\hat H_I$ in Eq.~\eqref{hint2}, with all $n_j=0$
and the form factors in Eq.~\eqref{bkq}.
Unfortunately, numerically exact solutions for this 
interacting problem 
are already out of reach except for very small system size. 
Here we instead proceed by employing the 
textbook Hartree-Fock (HF) approximation \cite{hftextbook}.  
However, going beyond this approximation is expected to cause at 
most quantitative -- but not qualitative -- modifications of  
the interaction-induced conductance discussed later on.
Note also that for the corresponding homogeneous ($d\to 0$) problem,
HF calculations give a good understanding of the physics away from
rational filling factors related to the fractional quantum Hall effect
\cite{rmp1,rmp2,nomura,moessner,joglekar,cote,christiane,faugeras}.
As HF parameters we choose the occupation numbers
\begin{equation}\label{hforder}
n_k = \left \langle c_{0,k}^\dagger c_{0,k}^{} \right \rangle, 
\end{equation}
where the expectation value is self-consistently taken with respect 
to the HF approximation of the zero-mode Hamiltonian.
For given filling factor $\nu$, 
the HF parameters have to be determined under the condition
$\sum_k n_k = N = \nu N_s.$
By choosing the $n_k$ as HF parameters, we disregard the possibility of
charge density wave or Wigner crystal formation \cite{hftextbook}.
Note that Wigner crystallization was reported in the
homogeneous case ($d=0$) with unscreened ($R\to \infty$) 
Coulomb interactions for certain filling factors $\nu$ \cite{joglekar,cote,christiane}.
However, for our MGW with externally screened interactions, 
see Eq.~\eqref{potential1}, we do not expect such phases.  

Defining single-particle energies as
\begin{equation} \label{hf2}
\varepsilon_{k}=  \sum_{k'} W_{k,k'} n_{k'} , \quad W_{k,k'}= 2
 W^{(0,0,0,0)}_{k,k';q=0},
\end{equation}
with the Coulomb matrix elements in Eq.~\eqref{cmt1}, 
the HF estimate for the ground-state energy reads
\begin{equation}  \label{ehf}
{\cal E}^{\rm HF}_0=\frac12 \sum_{k} \varepsilon_{k} n_k.
\end{equation}
The HF iteration starts with a normalized random initial
distribution for $n_k$, where we assume $n_{-k}=n_k$.
Next the HF energies $\varepsilon_k$ in
Eq.~\eqref{hf2} are computed, 
where $\varepsilon_{-k}=\varepsilon_k$ by virtue of 
the symmetry relations \eqref{symrel}.  The updated distribution $n_k$, which
is obtained by occupying the $N$ energetically lowest states, therefore always
remains even in $k$.  The scheme is then iterated until convergence has 
been reached.  

We find that the HF  ground-state energy converges quickly from above.
However, there are many local energy minima in occupation number space, 
and depending on the initial configuration one may converge to 
states of widely different energy.  We obtain the
global minimum by comparing converged results for
sufficiently many (typically a few hundred) randomly chosen initial states.  
The smooth behavior of all calculated quantities, such as the ground-state
energy or the effective Fermi momentum, on the system parameters
also confirms that this procedure reliably finds the HF ground state.
For the results shown here, graphene's fine structure
constant [see Eq.~\eqref{potential1}] was taken as $\alpha=0.5$, 
with the MGW width $d=2 l_B$ as in Fig.~\ref{fig2}.
To check our conclusions, we have performed additional calculations for 
other parameters, where the results (not shown) confirm the physical
picture presented in what follows.

\begin{figure}
\centering
\includegraphics[trim = 0 10 80 80,clip,width=7cm]{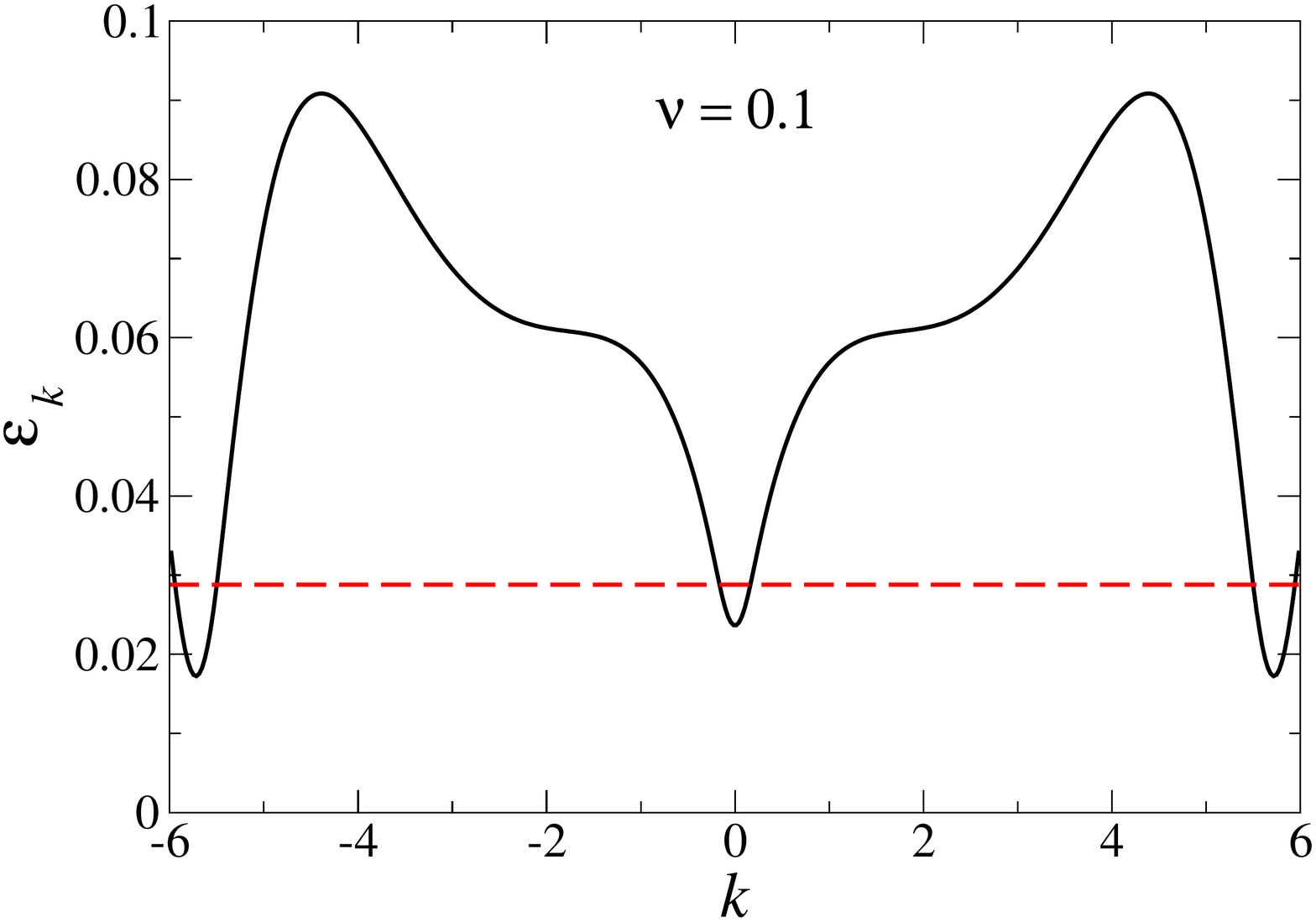}  


\includegraphics[trim = 0 10 80 80,clip,width=7cm]{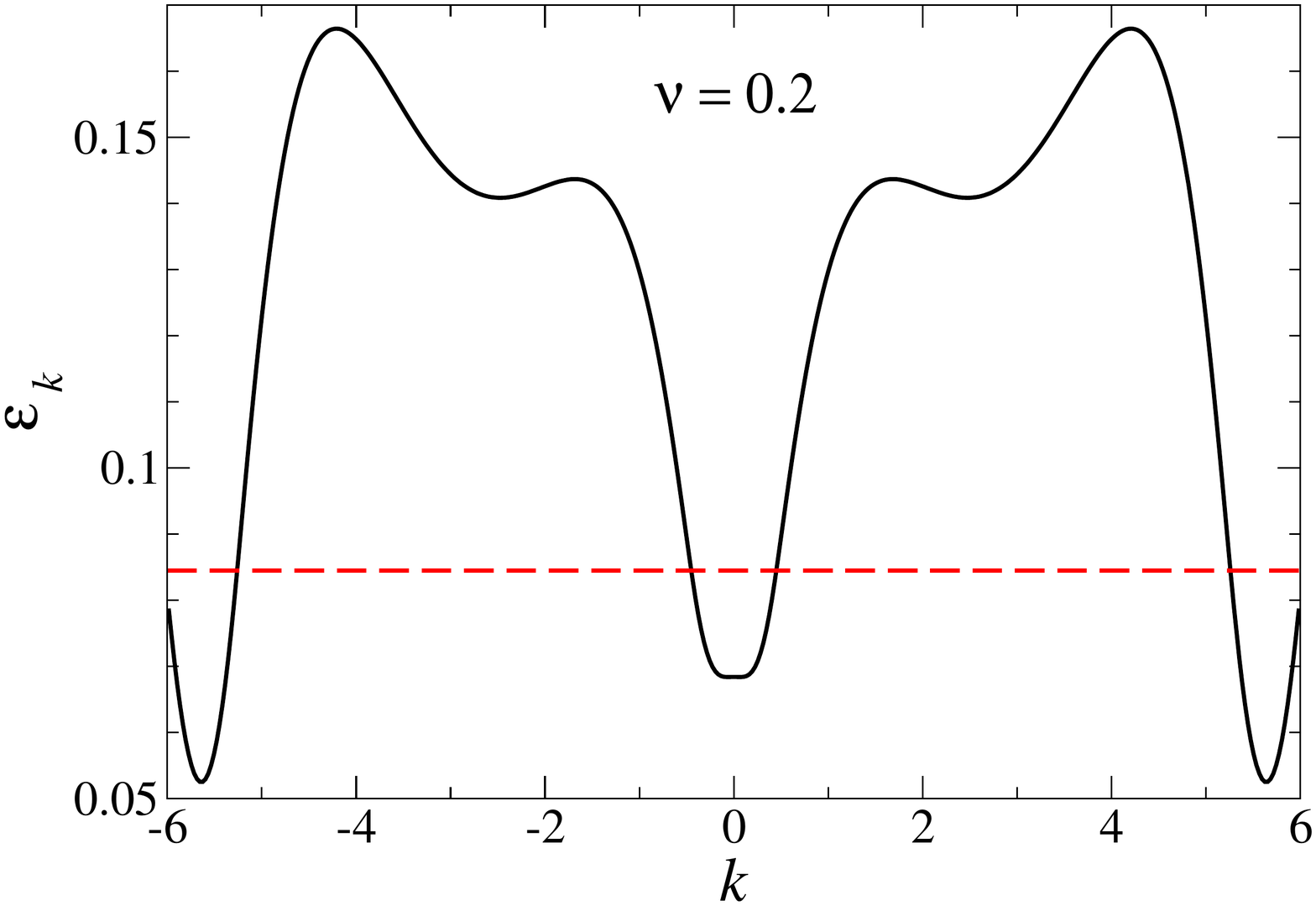}


\includegraphics[trim = 0 10 80 80,clip,width=7cm]{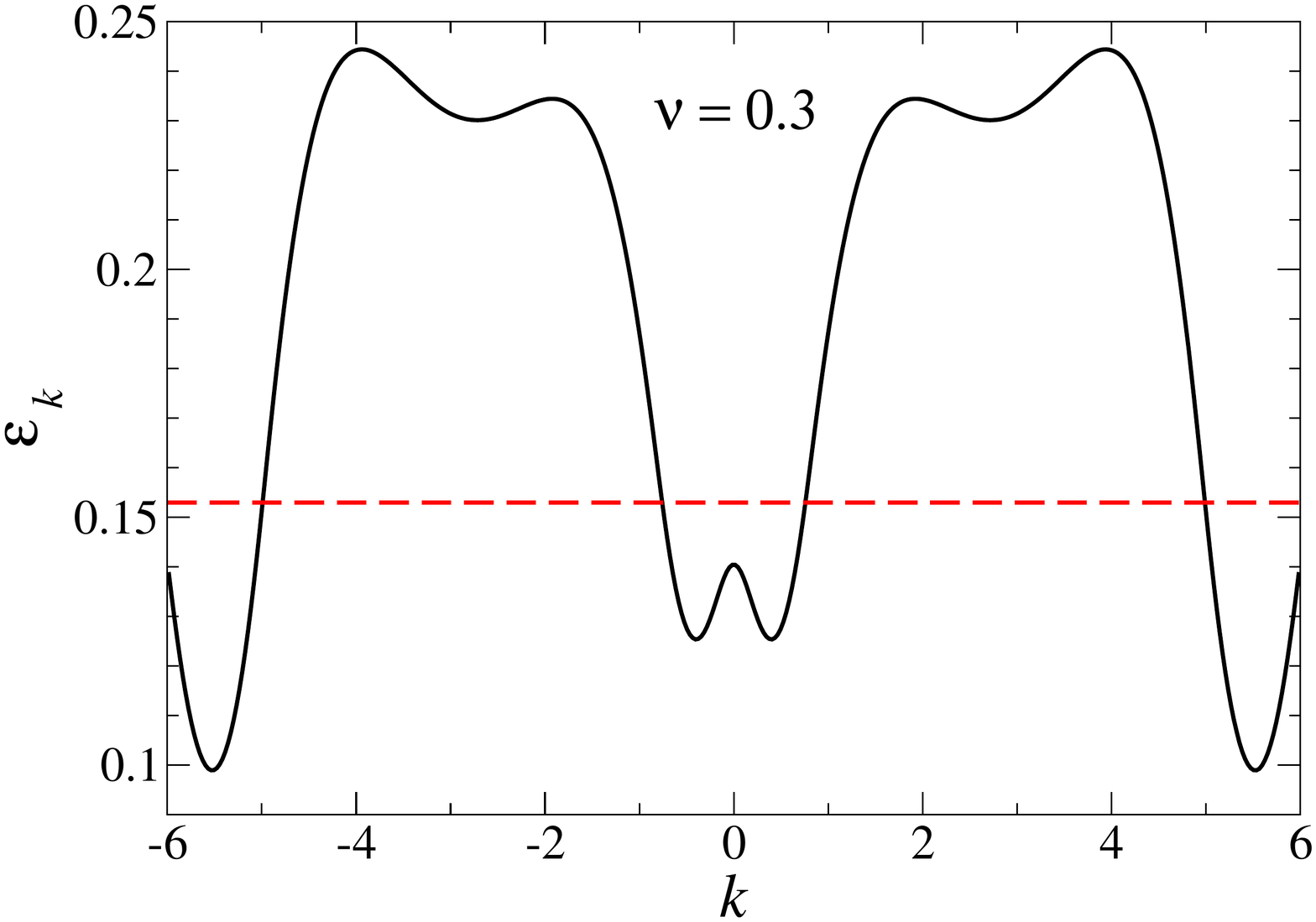}  


\includegraphics[trim = 0 10 80 80,clip,width=7cm]{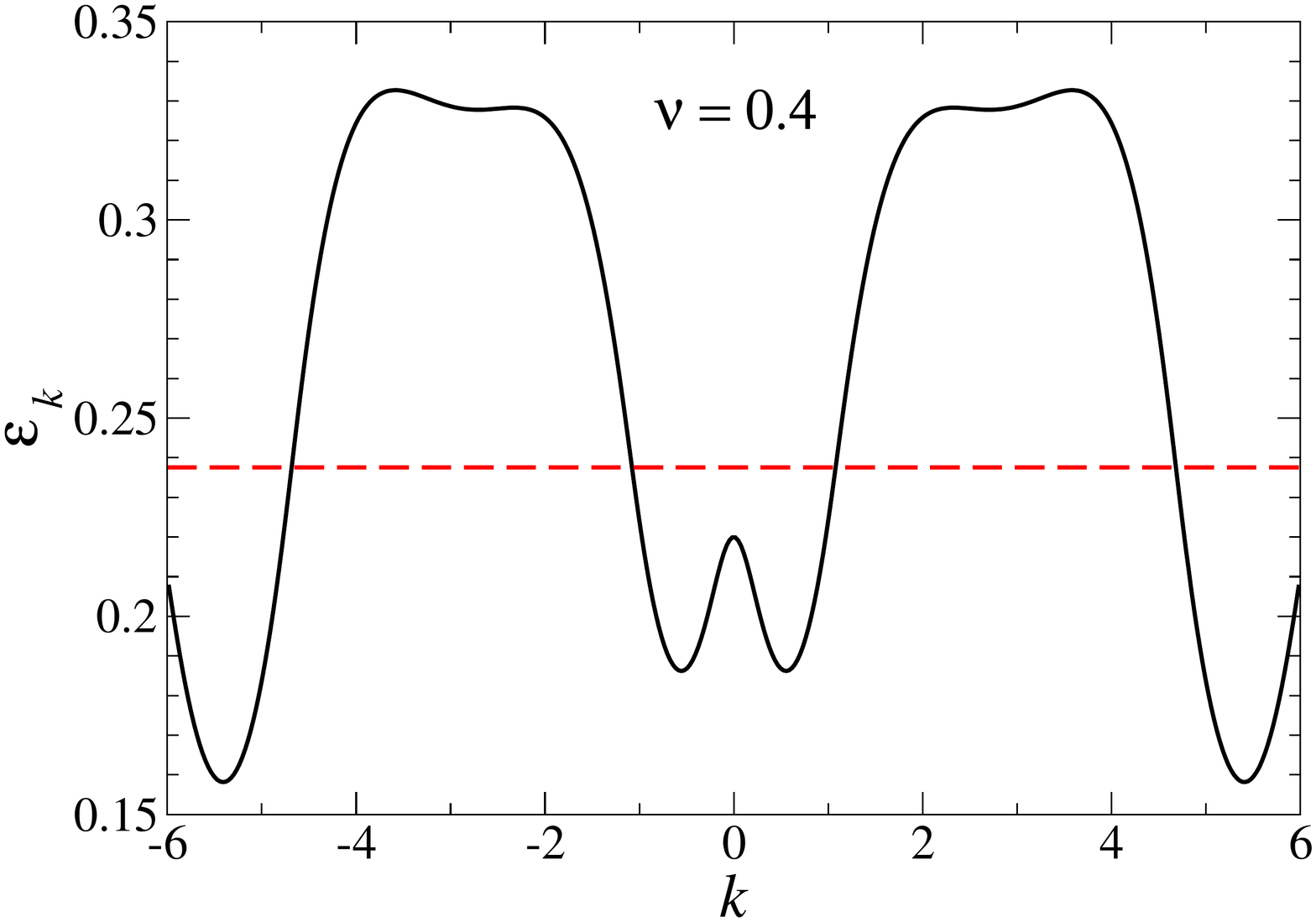}

\caption{\label{fig4}   
Self-consistent zero-mode HF energies $\varepsilon_k$ 
(in units of $E_B$) for various filling factors $\nu$. 
The red dashed horizontal lines denote the chemical potential at the 
respective filling, with all lower states occupied. 
We consider $\alpha=0.5$, $d=2l_B$, $R=d$, $L_x=16l_B$, 
and $N_s=400$ basis states, see
Eq.~\eqref{nsdef}. This parameter set corresponds to $L_y=209.4~l_B$,
where $|k|= \pi N_s/L_y = 6 l_B^{-1}$ represents the zone boundary.
}
\end{figure}

The zero-mode HF dispersion relation  is shown in Fig.~\ref{fig4}. 
For all studied filling factors, we found that the HF energies $\varepsilon_k$ 
develop a pronounced dip for small momenta, corresponding to 
states located mainly within the waveguide region, $|k|\alt d/l_B^2$.  
For larger $|k|$, we instead expect an almost flat dispersion 
(see below).  However, Fig.~\ref{fig4} reveals a preferential 
population of states at large momenta, which are spatially localized
near the boundaries at $x=\pm L_x/2$.
To understand this feature, it is instructive to briefly
study the homogeneous field case, $d\to 0$.  
Using Eqs.~\eqref{d0zmC} and \eqref{hf2}, we observe that 
for a $k$-independent distribution, $n_k=\nu$, the HF 
energies (\ref{hf2}) are given by
$\varepsilon_k\simeq (\nu L_y/2\pi) \int_{-L_x/2}^{L_x/2} 
dk' W_{k,k'}$.
Ignoring boundary effects, one could then effectively shift the 
integration variable $k'$ to absorb $k$, resulting in $k$-independent 
energies $\varepsilon_k$.  However, for states localized near the 
edge of the sample, dispersion is already predicted by this $d=0$ result.  
Such ``Pauli holes'' near the sample edges are clearly observed 
in Fig.~\ref{fig4}. However, these boundary states play no role for the 
zero-mode conductance $G$, since they have no significant interaction 
matrix elements with snake states. 
We can therefore safely ignore large-momentum states.  
In practice, we keep only single-particle states with $|k|\le k_c$,
where the momentum cutoff is chosen as $k_c\approx d/l_B^2$, 
cf.~Sec.~\ref{sec4c}.

For very small filling factor, $\nu \alt 0.08$, the $k=0$ minimum in 
the dispersion is above the Fermi level such that no small-$k$ modes 
are occupied. 
At larger fillings, however, Fig.~\ref{fig4} shows that this 
minimum in $\varepsilon_k$ drops below the Fermi level, and 
then evolves into a double minimum with increasing $\nu$. 
The latter feature can be understood from the reduced probability 
density $|\psi_{0,k}(x)|^2$ inside the waveguide region. This 
density has a local minimum at $x=kl_B^2$ (for $|k|<d/2l_B^2$) and thus 
comes with a reduced Coulomb repulsion cost.  Finally, for $\nu\agt 0.5$, 
we find that the HF energies may exceed the 
single-particle gap, $\varepsilon_k>E_{g}$. 
Since in that case our assumption of well-separated bands may break down, we
shall focus on the window $0.1\alt\nu\alt 0.4$ in what follows. 
Within this window, the perturbative approach in Sec.~\ref{sec4} is justified.

\begin{figure}
\centering
\includegraphics[trim = 0 10 80 80,clip,width=8cm]{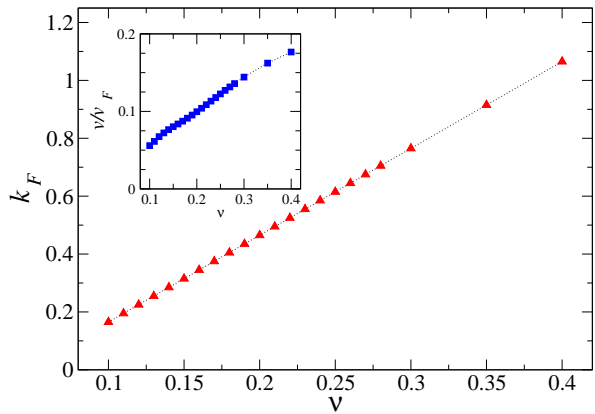}
\caption{\label{fig5}
Interaction-induced Fermi momentum $k_F$ vs filling factor $\nu$ for the
parameters in Fig.~\ref{fig4}. The dotted curve is a guide to the
eye only.  Note the approximately linear $\nu$-dependence.
The inset shows the effective Fermi velocity $v$ vs $\nu$, where   
$v$ is given in units of graphene's Fermi velocity $v_F$.  } 
\end{figure}

The respective Fermi level then intersects $\varepsilon_k$ for $k=\pm k_F$
with $k_F\alt d$.  The effective Fermi momentum, $k_F(\nu)$, 
and effective Fermi velocity, $v(\nu)$, 
are depicted as a function of the filling factor 
in Fig.~\ref{fig5}, where $v$ was obtained by linearizing
 $\varepsilon_k$ around $k=k_F$.  It is worth mentioning that while 
the results in Figs.~\ref{fig4} and \ref{fig5} were obtained for  
$\alpha= \alpha_0=0.5$, the corresponding results 
for $\alpha\ne \alpha_0$ follow from a simple scaling argument.  In particular, 
while $k_F$ is independent of $\alpha$, we find
$\varepsilon_k^{(\alpha)}=(\alpha/\alpha_0)\varepsilon_k^{(\alpha_0)}$ and thus
 $v^{(\alpha)} = (\alpha/\alpha_0) v^{(\alpha_0)}$.

The converged HF results for $\varepsilon_k$ can tentatively be 
interpreted as signature for an interaction-induced 
single-particle dispersion, where the low-energy physics is governed by a 
single pair of right- and left-movers with nearly linear 
dispersion relation and well-defined Fermi momentum.
Nonetheless, the conductance remains zero unless we also 
include virtual band transitions to $n\ne 0$ states. 
This statement holds true even under an exact treatment of interactions
(beyond HF theory), since no upper spinor components in Eq.~(\ref{zeropsi}) 
and thus no finite current matrix elements can then be generated. 

\section{Conductance} \label{sec4}

In this section, we study the zero-temperature linear 
{\sc DC-}conductance, $G$, of the MGW when the $n=0$ band is partially filled.
Within the HF approach in Sec.~\ref{sec3}, intra-band $n=0$ 
Coulomb interactions were shown to be responsible for an effective Fermi 
momentum $k_F=k_F(\nu)$, where $\nu$ is the filling factor, 
such that all single-particle states below the Fermi energy 
$E_F= \varepsilon_{k_F}$ are occupied.  
We here address the question: Are Coulomb
interactions able to induce a finite conductance in the
clean system?  Anticipating the affirmative answer to this question,
this feature offers a powerful novel way to directly probe 
electron-electron interaction effects in clean graphene samples
through transport measurements.

\subsection{Kubo formula}\label{sec4a}

We follow the standard Kubo linear-response formalism \cite{altland}, 
expressing the linear conductance as
\begin{equation}\label{kubo}
G = - e^2 \lim_{\omega\to 0} \frac{{\rm Im} \Pi^R(\omega)}{\omega},
\end{equation}
where $\Pi^R(\omega)$ is the Fourier transform of the 
retarded current-current correlation function,
\begin{equation}\label{curr-curr}
\Pi^R(t) = -i\Theta(t) C(t),\quad 
C(t)=\langle \Phi| [ \hat I(t), \hat I(0) ]_-|\Phi \rangle,
\end{equation}
with $\hat I(t)=e^{i Ht} \hat I e^{-i Ht}$, the 
Heaviside step function $\Theta(t)$, and
the normalized ground state $|\Phi\rangle$ of the 
full Hamiltonian $H$.  In second-quantized notation, 
the particle current along the $y$-direction is described by the operator 
\begin{equation}\label{currentdef}
\hat I = \sum_{n,n'} \sum_{k,k'} I_{n,k;n',k'} c_{n,k}^\dagger c_{n',k'}^{},
\end{equation}
with the matrix elements in Eq.~\eqref{currmatrel}.
Following a sequence of standard steps, the Fourier transform, 
$\tilde C(\omega)$, of the current-current 
correlator, $C(t)$, is related to the imaginary part of $\Pi^R(\omega)$, 
and thus represents a spectral function for current fluctuations.  Indeed, 
noting that $C^*(-t)=C(t)$ implies real-valuedness of $\tilde C(\omega)$, 
we find ${\rm Im}\Pi^R(\omega) = - \tilde C(\omega)/2$.
Furthermore, we note that $\tilde C(0)=0$ because of $C(-t)=-C(t)$.
The conductance thus follows as
\begin{equation}\label{kubo2}
G = \pi G_0 \frac{d\tilde C}{d\omega}(\omega=0),
\end{equation}
where $G_0=e^2/h$ is the conductance quantum. Writing $C(t)=X(t)-X(-t)$, we then 
need to evaluate the correlation function
\begin{equation}\label{xdef}
X(t) = \langle \Phi| \hat I(t) \hat I(0)|\Phi \rangle.
\end{equation}

At this point, it is useful to write the full Hamiltonian as
$H= \hat H_0 +\hat W$, where $\hat H_0$ captures not only the 
noninteracting part, cf.~Eq.~\eqref{h0}, but also includes 
the $n=0$ HF Coulomb interaction terms discussed in
Sec.~\ref{sec3}.  Writing
$\hat H_0 = \sum_{n,k}\tilde E_{n,k} c_{n,k}^\dagger c_{n,k}^{},$
with the effective single-particle energies 
$\tilde E_{n\ne 0,k} = E_{n,k}$ and $\tilde E_{0,k}=\varepsilon_k$,
all remaining interactions processes are then encoded by $\hat W$,
which in particular describes inter-band transitions.
For $\hat W=0$, the ground state $|\Phi\rangle=|\Phi_0\rangle$
corresponds to a Fermi sea with the occupation numbers 
\begin{equation}\label{fnk}
f_{n,k}= f(\tilde E_{n,k}),\quad f(E)= \Theta(E_F-E),
\end{equation}
where $f(E)$ is the Fermi function taken, for simplicity, at
zero temperature.
 
In order to include $\hat W$, it is convenient to evaluate $X(t)$ in Eq.~\eqref{xdef}
by using the Keldysh Green's function technique, where 
the time evolution proceeds from $t=-\infty$ to $t=+\infty$ (forward 
branch, $s=+$) and back from $t=+\infty$ to $t=-\infty$ (backward branch,
$s=-$) \cite{altland}.  From now on, we use the interaction picture, 
where time-dependent
operators are denoted by $\hat I(t)=e^{it\hat H_0} \hat I e^{-it \hat H_0}$.
We then have to double all dynamical fields according to the 
branch $s=\pm$ of the Keldysh contour, $\hat I(t)\to \hat I_s(t)$ and so on.
As a result, $X(t)$ takes the form 
\begin{equation}\label{pertstart}
X(t) =  \left\langle \Phi_0 \left|
{\cal T}_C \left[ S(\infty) \hat I_-(t)\hat I_+(0)\right] \right|\Phi_0\right\rangle,
\end{equation}
where ${\cal T}_C$ is the time-ordering operator along the
Keldysh contour, and the time-evolution operator reads
\begin{equation}\label{evolutionop}
S(\infty) = {\cal T}_C \exp\left(-i\int_{-\infty}^\infty d\tau
\sum_{s=\pm} s \hat W_s(\tau)  \right).
\end{equation}

\subsection{Diagrammatic expansion}\label{sec4b}

\begin{figure}
\centering
\includegraphics[width=6cm]{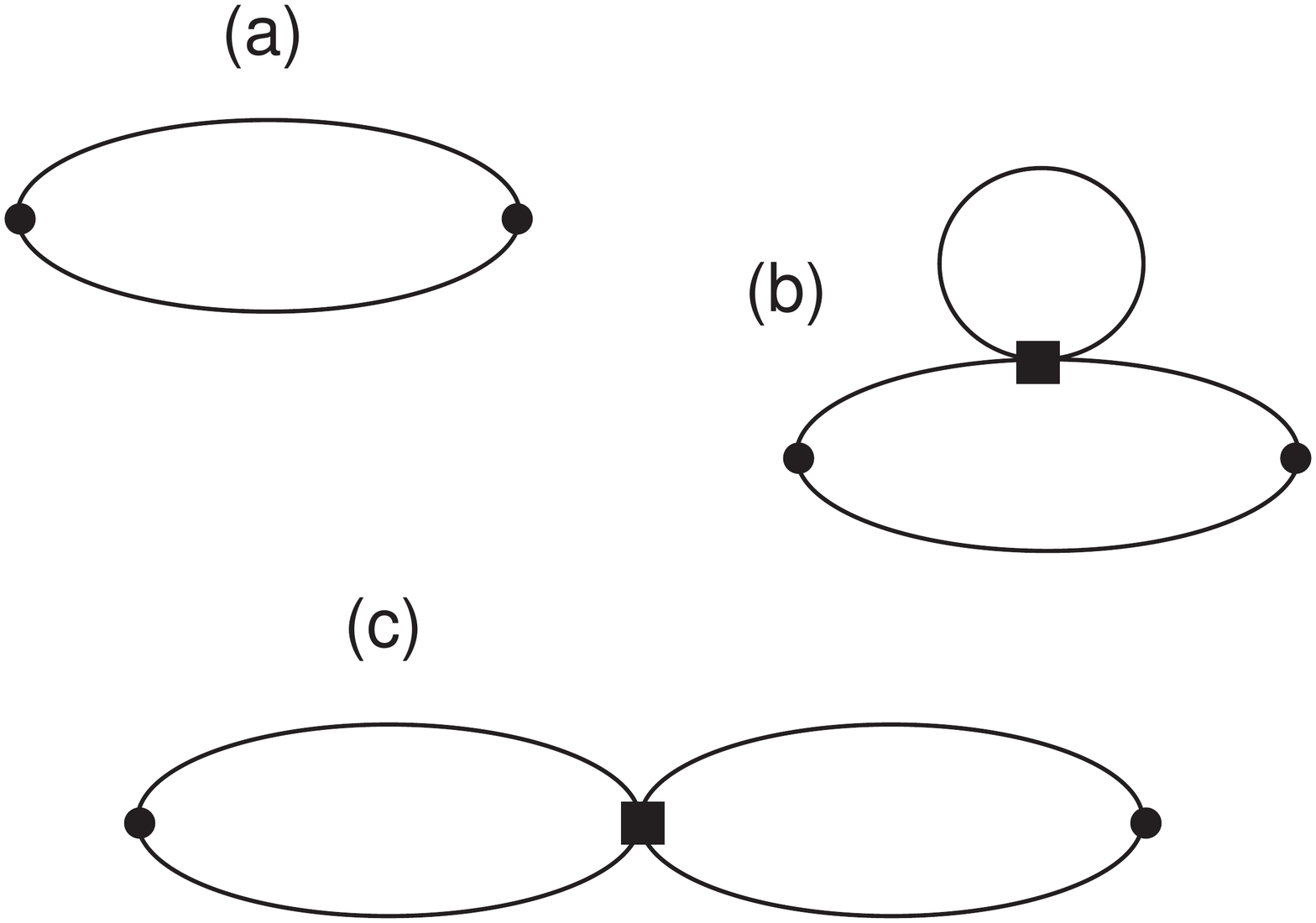}
\caption{\label{fig6}  Diagrammatic expansion of the conductance 
(see main text).  Panel (a) shows the zeroth-order 
($m=0$) diagram, while (b) and (c) 
refer to the two first-order ($m=1$) diagrams.  }
\end{figure}

\begin{figure}
\centering
\includegraphics[width=7cm]{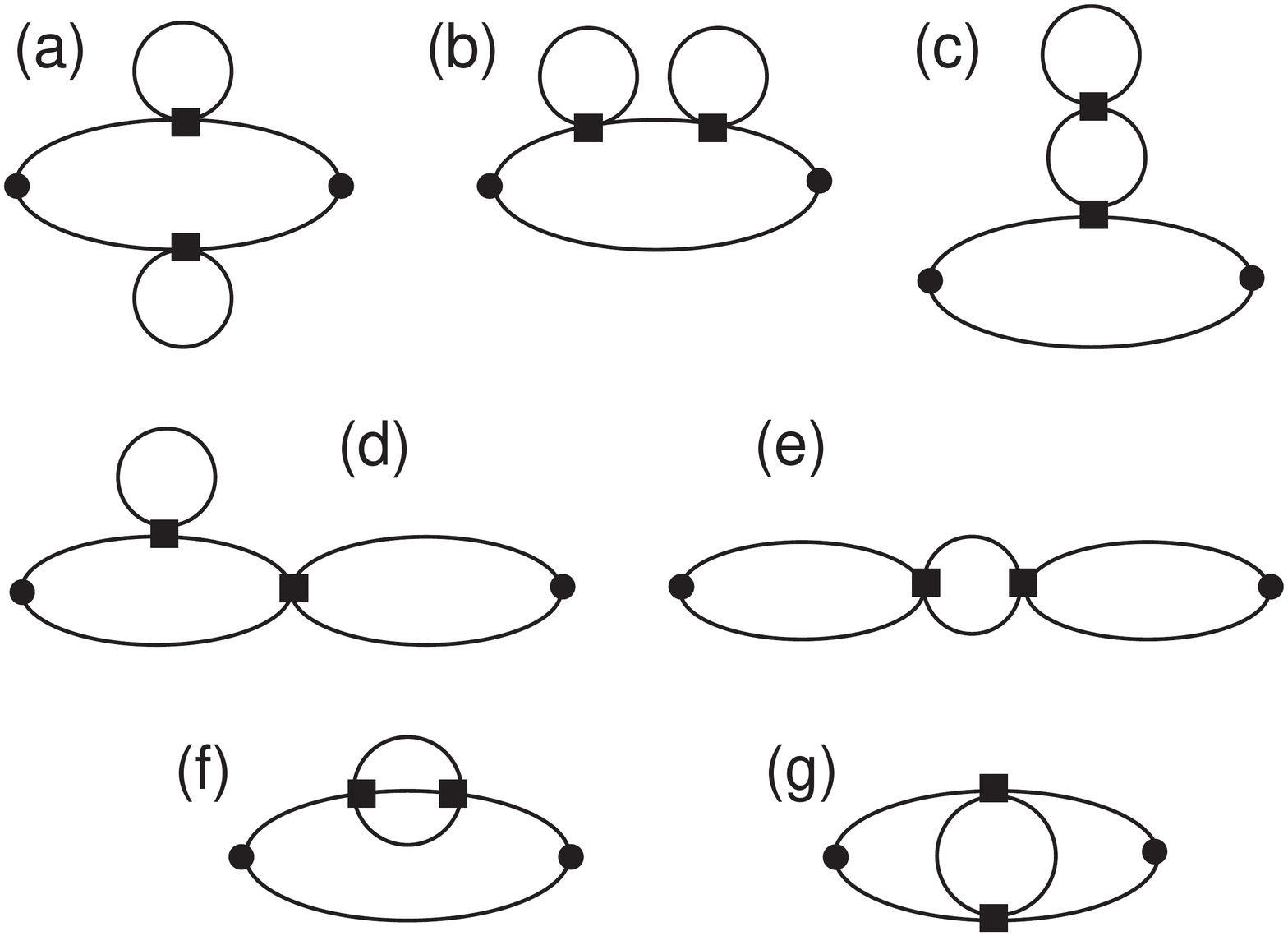}
\caption{\label{fig7}  Same as Fig.~\ref{fig6} but for the 
second-order ($m=2$) diagrams.  }
\end{figure}

Our strategy will be to compute the conductance as perturbation
series in the interaction term $\hat W$, which captures the effects of virtual
band transitions.  Going up to second order in $\hat W$ yields
\begin{equation}\label{condexpansion}
G = G^{(0)}+ G^{(1)}+ G^{(2)} + O(\hat W^3).
\end{equation}
Expanding Eq.~\eqref{evolutionop}
in powers of $\hat W$, we obtain 
a corresponding series for 
$X(t)=X^{(0)}+X^{(1)}+X^{(2)}+\cdots$,
with the $m$-th order term given by
\begin{eqnarray}\nonumber
&& X^{(m)}(t) = \frac{(-i)^m}{m!}\int_{-\infty}^\infty
d\tau_1\cdots d\tau_m \sum_{s_1,\ldots,s_m=\pm} s_1\cdots s_m \\
&&\times \left\langle \Phi_0 \left|
{\cal T}_C \left[ \hat W_{s_1}(\tau_1) \cdots \hat W_{s_m}(\tau_m) 
\hat I_-(t) \hat I_+(0)\right] \right|\Phi_0\right\rangle.
\label{expastart}
\end{eqnarray}
Application of Wick's theorem to the time-ordered products of
noninteracting fermion operators in Eq.~\eqref{expastart} 
allows one to interpret such expressions in a diagrammatic language.
The propagator (``line'') in a given diagram then
corresponds to the Keldysh Green's function (with Keldysh indices 
$s,s'=\pm$) \cite{altland},
\begin{equation}\label{gf1}
{\cal G}^{(s,s')}_{n,k}(t-t') = -i\langle \Phi_0|{\cal T}_C[ c_{n,k,s}^{}(t)
 c^\dagger_{n,k,s'}(t')]|\Phi_0\rangle,
\end{equation}
with the Fourier-transformed components
\begin{eqnarray}\nonumber
\tilde {\cal G}_{n,k}^{(s,s)}(E) &=& \frac{s}{E-\tilde E_{n,k} +i s\ 
{\rm sgn}\left(\tilde E_{n,k}-E_F\right) \ 0^+ } ,   \\ \label{gkf}
\tilde {\cal G}_{n,k}^{(s,-s)} (E) &=& 2\pi i s f(sE) \ 
\delta\left(E-\tilde E_{n,k} \right),
\end{eqnarray}
where $f(E)$ is the Fermi function, see Eq.~\eqref{fnk}.
All conductance diagrams up to order $m=2$ are shown in 
Figs.~\ref{fig6} and \ref{fig7}. They are constructed from the following rules:
\begin{itemize}
\item Each diagram must contain two external (two-point) vertices 
representing the current operators $\hat I_-(t)$ and $\hat I_+(0)$. These vertices carry 
the respective Keldysh index $s=\pm$, and are
denoted by filled circles in Figs.~\ref{fig6} and \ref{fig7}. 
\item For an $m$-th order diagram, there are $m$ internal (four-point) 
vertices representing the interaction Hamiltonian, i.e., factors 
$s\hat W_{s}(\tau)$. These vertices are denoted by filled squares.  
\item Using the Keldysh normalization condition 
$\langle \Phi_0|S(\infty)|\Phi_0\rangle=1$ as well as
$\langle \Phi|\hat I_\pm (t)|\Phi\rangle=0$, 
only connected diagrams have to be taken into account.
\end{itemize}
We mention in passing that in evaluating low-order diagrams 
it is useful to exploit a ``selection rule'' that
allows one to discard certain contributions
without detailed calculation. Indeed, owing to the vanishing current 
matrix elements in Eq.~\eqref{zerocur},
a diagrammatic contribution is zero whenever a current vertex with 
Keldysh index $s=\pm$ is connected to two
other vertices that both have the opposite Keldysh index $-s$ \cite{foot}. 

At this stage, it is convenient to introduce several auxiliary matrices.  
First, we define the current matrix $I$   
through its matrix elements $I_{n,k;n',k'}$ in 
Eq.~\eqref{currmatrel}. Second, the hybridization matrix $\Lambda$ with 
\begin{equation}\label{hybmat}
\Lambda_{n,k;n',k'}=  \frac{\delta_{k,k'}(1-\delta_{n,n'})}
{\tilde E_{n,k}- \tilde E_{n',k}} 
\sum_{n_1,k_1} f_{n_1,k_1} W_{k,k_1;q=0}^{(n,n_1,n_1,n')}
\end{equation}
describes interaction-induced virtual transitions between 
different bands.
Here the intermediate summation comes from a fermion loop.  
We notice that $\Lambda$ is diagonal in $k$-space 
and hence can indeed be interpreted as hybridization matrix.
The commutator of the above matrices is given by 
\begin{equation}\label{deltai}
\delta I = [I,\Lambda]_-,
\end{equation}
which captures the current matrix renormalization by
virtual interband transitions to leading order. 
Finally, we introduce a fluctuation matrix $K$, with matrix elements
\begin{eqnarray}
K_{n,k;n'k'} & = &  \sum_{n_1,k_1;n_2,k_2}
\delta_{k_2,k_1+k'-k} \frac{ f_{n_1,k_1}-f_{n_2,k_2} }{
\tilde E_{n_2,k_2}-\tilde E_{n_1,k_1}}
 \nonumber \\ &\times& I_{n_1,k_1;n_2,k_2}  W^{(n_2,n,n_1,n')}_{k_2,k;k-k_1}.
\label{kdef}
\end{eqnarray}
Below, the ``$\circ$'' symbol denotes Fermi level convolution, i.e., the
matrix elements of $A\circ B$ are given by
\begin{equation}\label{aux1}
(A\circ B)_{n,k;n',k'} = \sum_{\pm} A_{n,k;0,\pm k_F} B_{0,\pm k_F; n',k'},
\end{equation}
and ``Tr$_F$'' denotes a Fermi level trace, 
\begin{equation}\label{aux2}
{\rm Tr}_F A = \sum_\pm A_{0,\pm k_F; 0,\pm k_F}.
\end{equation}

Let us now start with the zeroth-order conductance contribution, 
where inter-band transitions are absent.  From the discussion 
in Sec.~\ref{sec3}, this term is expected to vanish identically, $G^{(0)}=0$.  
There is only a single $m=0$ diagram, represented by the polarization 
bubble in Fig.~\ref{fig6}(a).
By virtue of Eqs.~\eqref{expastart} and \eqref{gf1}, this diagram leads
to the correlation function 
\begin{equation}
X^{(0)}(t) = \sum_{n,k;n',k'} 
\left| I_{n,k;n',k'}\right|^2 {\cal G}_{n,k}^{(-,+)}(t) 
{\cal G}^{(+,-)}_{n',k'}(-t).
\end{equation}
Performing a Fourier transformation, we find 
\begin{eqnarray}
\tilde X^{(0)}(\omega) &=& 2\pi \sum_{n,k;n',k'} 
\left| I_{n,k;n',k'}\right|^2 \\ \nonumber
&\times&  (1-f_{n,k}) f_{n',k'} \delta(\tilde E_{n,k}
-\tilde E_{n',k'}-\omega),
\end{eqnarray}
and with Eq.~\eqref{kubo2}, the respective zero-temperature
conductance contribution is
\begin{equation}\label{zerothg}
G^{(0)}/G_0 =  \frac{\pi^2\rho_0^2}{2}  \ {\rm Tr}_F  (I\circ I).
\end{equation}
Here $\rho_0 = 1/\pi v$ denotes the density of states, 
 with $v=\partial_{k}\varepsilon_{k=k_F}$, and the symbols ``$\circ$'' and ``Tr$_F$''
were defined in Eqs.~\eqref{aux1} and \eqref{aux2}, respectively.
Since all current matrix elements appearing in Eq.~\eqref{zerothg} vanish
by virtue of Eq.~\eqref{zerocur}, this calculation confirms the expected 
result $G^{(0)}=0$.

Let us then turn to the first-order ($m=1$) diagrams. After some algebra, 
 the corresponding conductance contribution takes the form 
\begin{equation}\label{correct1}
G^{(1)}/G_0 = -2\pi^2 \rho_0^2 \ {\rm Tr}_F \left[ I\circ (\delta I - K) \right],
\end{equation}
where the $\delta I$-term comes from diagram (b) and the $K$-term from
diagram (c) in Fig.~\ref{fig6}, respectively, see also Eqs.~\eqref{deltai} and 
\eqref{kdef}.
Because of the vanishing zero-mode current matrix elements
in Eq.~\eqref{zerocur}, also this conductance contribution
vanishes, $G^{(1)}=0$.  
However, it is worth mentioning that for a conventional system with non-zero
current matrix elements, e.g., the finite-energy bands for our MGW,
Eqs.~\eqref{zerothg} and \eqref{correct1} yield finite results,
corresponding to the ``noninteracting'' conductance 
and the ballistic version of the ``interaction correction'' 
\cite{aa,zala,kupfer}, respectively.

In order to encounter a finite conductance in our zero-mode system, 
we have to go up to second order ($m=2$).
All topologically distinct $m=2$ diagrams are shown in Fig.~\ref{fig7}.  
Using the above selection rule, we find that diagram 
(c) also gives no conductance contribution.  Moreover, diagrams 
(f) and (g) vanish as well since
they involve products of more than two Fermi factors which never 
satisfy the resulting energy constraints. 
Within second-order perturbation theory, the conductance is thus obtained 
from the remaining diagrams in Fig.~\ref{fig7},
$G=G^{(2)}=G_a+G_b+G_d+G_e$, which yield a finite result.  
Evaluating diagrams (a) and (b) together gives 
\begin{equation}
(G_a+G_b)/G_0 = 2\pi^2 \rho_0^2 \ {\rm Tr}_F (\delta I \circ \delta I).
\end{equation}
Similarly, diagram (d) yields 
\begin{equation}
G_d/G_0 = -4\pi^2 \rho_0^2 \ {\rm Tr}_F (\delta I \circ K),
\end{equation}
while diagram (e) produces the contribution 
\begin{equation}
G_e/G_0 = 2\pi^2\rho_0^2 \ {\rm Tr}_F (K\circ K).
\end{equation}
Collecting all diagrams, we obtain a manifestly positive and general result 
for the zero-mode conductance,
\begin{equation}\label{finalcond}
G/G_0=2\pi^2 \rho_0^2 \ {\rm Tr}_F\left[ (\delta I-K)\circ (\delta I-K) \right].
\end{equation}
We note in passing that Eq.~\eqref{finalcond} does 
not apply for $d\to 0$, where the zero-mode dispersion 
$\varepsilon_k$ becomes flat and hence $k_F$ is not defined anymore.   

\subsection{Zero-mode MGW conductance} \label{sec4c}

We next discuss the zero-mode MGW conductance predicted by Eq.~\eqref{finalcond},
adopting the parameters in Sec.~\ref{sec3}.  Our main goals 
are (i) to reliably demonstrate the existence of a finite zero-mode 
conductance, and (ii) to clarify its filling dependence.
In order to simplify the numerical evaluation, 
which is quite cumbersome due to the presence of 
interaction matrix elements connecting all different bands,
we shall here evaluate Eq.~\eqref{finalcond} by taking into account only the 
three bands $n=-1,0,1$ sketched in Fig.~\ref{fig2}.  
Indeed, the $n=\pm 1$ bands are energetically closest to 
the $n=0$ modes and therefore produce the main conductance contribution.
Moreover, to avoid spurious finite-size effects,
we introduce a momentum bandwidth $k_c$ restricting 
the single-particle Hilbert space to states with $|k|\le k_c$, see
Sec.~\ref{sec3}. For the results below, where $d=2l_B$, 
we chose the momentum cutoff $k_c =1.6 l_B^{-1}$.  
However, taking other values within the range 
$1.5\alt k_c l_B\alt 1.7$ also gave essentially identical results.

\begin{figure}
\centering
\includegraphics[trim = 0 10 80 80,clip,width=8cm]{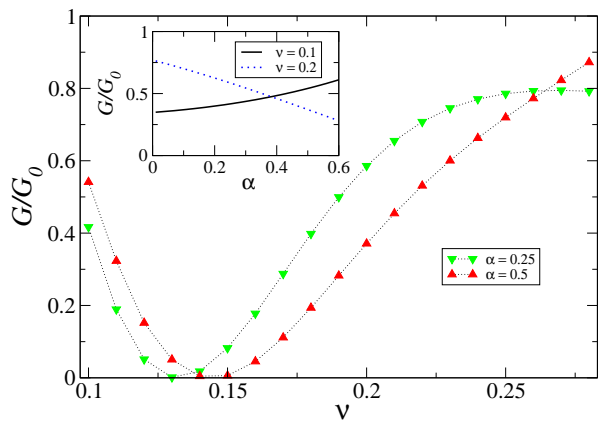}
\caption{\label{fig8} 
Zero-mode waveguide conductance $G$ (in units of $G_0=e^2/h$) 
 vs filling $\nu$, see Eq.~\eqref{finalcond},
for two values of the fine structure constant $\alpha$.
Dotted curves are guides to the eye only.  
All other parameters are as in Fig.~\ref{fig4}. The inset shows
the $\alpha$-dependence at two selected fillings.  
}
\end{figure}

The resulting zero-temperature conductance is illustrated in Fig.~\ref{fig8}. 
The main panel shows that the conductance strongly depends on the 
zero-mode filling factor $\nu$, with a pronounced minimum 
around $\nu=\nu_{\rm min}$, 
where we find $\nu_{\rm min}\approx 0.145$ at $\alpha=0.5$.  
Near this minimum, $G$ becomes very small. 
The existence of the minimum in $G(\nu)$ can be rationalized by 
analyzing the Coulomb-assisted hybridization of zero modes with 
the $n=\pm 1$ bands.  For small filling $\nu$, and 
therefore small chemical potential, see Sec.~\ref{sec3},
the band $n=-1$ still remains close to the $n=0$ states.
In that case, the current renormalization effects encoded in
$\delta I$ are found to dominate the conductance in Eq.~\eqref{finalcond}.  
With increasing filling, this conductance contribution begins to 
weaken, while the Fermi level gets closer and closer to the $n=1$ band. 
Eventually, at large filling, the 
fluctuation matrix $K$ instead dominates in Eq.~\eqref{finalcond}.
We then encounter a nearly perfect cancellation of the
$\delta I$ and $K$ terms for filling factor $\nu\approx \nu_{\rm min}$.

The inset of Fig.~\ref{fig8} shows the $\alpha$-dependence of the conductance
for two fillings $\nu$, chosen below and above $\nu_{\rm min}$, respectively. 
Two comments are in order here. First, the limit $\alpha\to 0$ seems to result in
a finite conductance. This may come as a surprise, since in the 
absence of interactions, we know that $G=0$ must hold. 
However, one should keep in mind that our HF approach in Sec.~\ref{sec3} 
applies only if the frequency scale $\omega$ at which the 
conductance is probed remains well below the effective bandwidth of 
the zero mode.  Since this $\varepsilon_k$ bandwidth is proportional to $\alpha$,
it vanishes in the limit $\alpha\to 0$, and 
the $\omega\to 0$ limit implicit in Eq.~\eqref{finalcond} 
cannot be taken anymore.
Second, the different $\alpha$-dependence at the two fillings observed in 
the inset of Fig.~\ref{fig8} simply reflects the fact that $\nu_{\rm min}$ 
increases with increasing $\alpha$, cf.~the main panel.
This shift of $\nu_{\rm min}$ in turn can be rationalized by noting that the 
chemical potential also moves up with increasing $\alpha$, and thus
 the cancellation point between the $\delta I$ and $K$ 
contributions is slightly shifted towards larger fillings. 

It is instructive to contrast the above results for the zero-mode 
conductance with 
the conductance found when the chemical potential intersects one of the 
$n\ne 0$ bands instead.  This case has been studied in detail in 
Ref.~\cite{hausler2}, where a completely different behavior has been reported.
When contacted by wide electrodes, the zero-temperature conductance 
then assumes a quantized value (in units of $G_0$),
where transport proceeds predominantly through snake states. 
This quantization can be rationalized by noting that 
right- and left-moving snake states are spatially separated \cite{hausler2}. 
In addition, no pronounced dependence of the conductance on the respective band
filling is expected, in marked contrast to the zero-mode case shown in 
Fig.~\ref{fig8}. As we elaborate further in Sec.~\ref{sec5}, the 
finite-energy bands correspond to a realization of the conventional 
TLL phase, which could also be detected through the predicted 
power-law corrections to the conductance at finite temperatures 
\cite{hausler2}.

To summarize, the strong dependence of the zero-mode conductance $G(\nu)$ 
on the filling factor $\nu$, see Fig.~\ref{fig8}, should allow for a
clear experimental signature of the predicted interaction-induced 
conducting phase.  
As we argue in the next section, such a state is distinct from a TLL state,
and hence also not described by Fermi liquid theory.

\section{Discussion} \label{sec5}

In strictly 1D band metals, it is well known that low-energy 
excitations near the Fermi points at $\pm k_F$ are severely
restricted as a consequence of phase-space limitations \cite{hftextbook},
and therefore any nonzero electron-electron interaction slightly destabilizes
the Fermi liquid \cite{gogolin-book}.  The resulting
phase is commonly coined ``Tomonaga-Luttinger liquid'' (TLL), 
where all low-energy properties of the system are fully determined by just two
parameters when disregarding the spin sector \cite{haldane,voit,giamarchi}. 
If Galilei invariance holds in addition, which 
is the case for the continuum model considered here, a single TLL
parameter, $g_{\rm TLL}$, remains. 
The Fermi liquid case is recovered for $g_{\rm TLL}=1$.
 The Kubo conductance of an infinitely long and clean TLL 
is $G=g_{\rm TLL}G_0$ \cite{apelrice}, and
single-particle correlation functions exhibit power-law 
behavior with exponents 
controlled by $g_{\rm TLL}$ \cite{gogolin-book,voit}.
To give just one example,  the equal-time
single-particle Green's function,
cf.~Eq.~\eqref{fieldop}, has the asymptotic power-law decay 
$\langle \hat\Psi(x,y)\hat\Psi^{\dagger}(x,0)\rangle
\sim |y|^{-\gamma}$
with $\gamma=1+[g_{\rm TLL}+1/g_{\rm TLL}-2]/4$. 
The value of $g_{\rm TLL}$ is fixed by the interaction strength through 
the ground-state compressibility \cite{voit},
\begin{equation}\label{gcompress}
g_{\rm TLL}=\left[\frac{\pi}{v}\frac{\partial^2({\cal E}_0/L_y)}{\partial
k_F^2}\right]^{-1/2},
\end{equation}
where ${\cal E}_0$ is the ground-state energy and
$v=|\partial_k\varepsilon_{k=k_F}|$ 
the single-particle velocity, see also Ref.~\cite{kecke}.  
Precisely this scenario has previously been identified in our
MGW for all $n\ne 0$ snake states \cite{hausler2}.  Without
interactions, snake states propagate uniformly at the Fermi
velocity $v_F$ of the graphene host.
They represent spatially separated chiral branches located
near either of the two parallel zero lines of the magnetic field.
The value of $g_{\rm TLL}$ is then governed by Coulomb
interactions between these oppositely moving branches and can be tuned directly 
via the MGW width $d$. 

On the other hand, the partially filled $n=0$ MGW band 
investigated here clearly does not fit into the above TLL framework. First,
without interactions there is no Fermi surface, since the zero-energy 
level is strictly flat.
Second, when accounting for intra-band Coulomb interactions on
the level of the HF approximation, we found an interaction-induced
dispersion, where the effective single-particle energies allow us 
to define  a Fermi momentum, $k_F$, and a Fermi velocity, $v$,
see Fig.~\ref{fig5}. One may then naively conclude
 that once again a TLL emerges, where $g_{\rm TLL}$ follows from
Eqs.~\eqref{ehf} and \eqref{gcompress}. 
However, this value of $g_{\rm TLL}$ does not describe correctly
the conductance of the system which is zero. 
We stress that the vanishing conductance of the $n=0$ band alone holds
true even when performing an exact calculation, see Sec.~\ref{sec3}.
Therefore, contrary to all $n \ne 0$ bands, we conclude 
that the $n=0$ band cannot be described by a bosonized Gaussian
field theory underlying the Luttinger liquid concept \cite{haldane}. 
Only when accounting for virtual interband transitions to conducting
$n \ne 0$ bands, the $n=0$ band acquires a nonzero conductance 
for which we find a quite peculiar dependence on the filling $\nu$.

We expect that apart from the MGW studied here, 
similar behavior should be observable also in other settings.  For instance,
consider metallic carbon nanotubes with a magnetic field applied perpendicular to the tube
axis \cite{tcnt}. This field should be inhomogeneous on the scale of 
the tube radius, such that a non-zero net magnetic flux $\Phi$ penetrates
the tube, since only then a finite degeneracy, $\Phi/(hc/e)$, of the 
$n=0$ Landau level is guaranteed by index theorems \cite{goerbig}, 
see also Ref.~\cite{prada}.  
Accounting for intra-Landau level interactions, 
we expect that dispersion of the $n=0$ level is created.
Nonetheless, the system will still exhibit insulating behavior when the Fermi
level is close to neutrality, and only when including
inter-Landau level interactions, a non-zero conductance can emerge. 
Similar to the case of the MGW, the actual 
 value of the conductance will then give direct information on the
strength of the Coulomb interaction, and 
the conductance behavior of the $n=0$ level 
should again significantly differ from all $n\ne 0$ bands.

Our theory assumes  that one works with a ballistic (disorder-free)
sample. We expect that very weak disorder will not qualitatively
change the scenario outlined above, but strong disorder will 
introduce localization and thereby destroy the 
physics described here.  Since ballistic transport is nowadays
reachable in high-quality graphene devices, our results should
be testable in the near future.

To conclude, the zero-energy levels in a clean magnetic graphene 
waveguide are predicted to display qualitatively different 
conductance features than the bands of nonzero energy. For the latter,  
the zero-temperature conductance is quantized in multiples of the 
conductance quantum. In contrast, the zero-mode 
conductance is non-universal with a strong dependence on the filling factor.
Since without Coulomb interactions this conductance vanishes 
identically, transport experiments offer a direct interaction probe.
We hope that our predictions can soon be put to an 
experimental test and will inspire further studies of this 
novel transport regime.

\acknowledgments

We thank A. De Martino, E. Eriksson, and S. Plugge 
for useful discussions.  W.H.~thanks P.~H\"anggi for long-lasting support.
This work was supported by the network SPP 1459 of the Deutsche 
Forschungsgemeinschaft (Bonn).

\appendix 

\section{Spectrum of MGW}\label{appa}

In this Appendix, we provide some details concerning Sec.~\ref{sec2a}.
For given $k$, using the natural units in Eq.~\eqref{units},
$H_0$ in Eq.~\eqref{h0} reduces to the 1D Hamiltonian
\begin{equation}
H_0^{(k)} = -i\sigma_x\partial_x+ [k+A(x)]\sigma_y,
\end{equation}
where $A(x)$ in Eq.~\eqref{vecpot} is antisymmetric 
under $x$-inversion, ${\cal R}_x: x\to -x$.
Due to this property, $H_0^{(k)}$ exhibits inversion symmetry,
\begin{equation}\label{conseta}
\left[ H^{(k)}_0, \Xi\right ]_- = 0, \quad 
\Xi = {\cal R}_k {\cal R}_x {\cal C},
\end{equation}
where ${\cal R}_{k}: k\to -k$ inverts $k$ and ${\cal C}$ 
denotes complex conjugation.  The operator $\Xi$ has 
eigenvalues $\xi=\pm$, and Eq.~\eqref{conseta} 
implies that the eigenstates in Eq.~(\ref{free-solution})
obey the relation
\begin{equation}\label{symrel0}
\Xi\psi_{n,k}(x) = \psi_{n,-k}^\ast(-x)= \pm \psi_{n, k}(x).
\end{equation}

Next, we note that for $x<-d/2$, the general solution at 
energy $E=E_{n,k}$ can be written in the form \cite{ademarti}
\begin{equation}\label{psiI}
\psi_{n,k;I} (x) = C^{(n,k)}_I \left ( \begin{array}{c} 
D_{-1+E^2/2} \left(-\sqrt{2}(x+k+d) \right) 
\\-\frac{i\sqrt{2}}{E} D_{E^2/2} \left(-\sqrt{2}(x+k+d) \right) 
\end{array} \right),
\end{equation}
with a coefficient $C_I^{(n,k)}$ and the parabolic cylinder
function $D_p(x)$ \cite{gradst,abramowitz}. 
Similarly, with coefficients $C^{(n,k)}_{\pm,II}$, 
the solution in the waveguide region $|x|<d/2$ reads
\begin{equation}\label{psiII}
\psi_{n,k;II} (x) = \sum_\pm C^{(n,k)}_{\pm,II}  \left ( \begin{array}{c} 
D_{E^2/2} \left(\pm \sqrt{2}(x-k) \right) 
\\ \mp \frac{iE}{\sqrt{2}} D_{-1+E^2/2} \left(\pm \sqrt{2}(x-k) \right) 
\end{array} \right),
\end{equation}
while for $x>d/2$, one finds
\begin{equation}\label{psiIII}
\psi_{n,k;III} (x) = C_{III}^{(n,k)} \left ( \begin{array}{c} 
D_{-1+E^2/2} \left(\sqrt{2}(x+k-d) \right) 
\\ \frac{i\sqrt{2}}{E} D_{E^2/2} \left(\sqrt{2}(x+k-d) \right) 
\end{array} \right).
\end{equation}
The symmetry relations in Eq.~\eqref{symrel0} now 
connect the coefficients in Eqs.~\eqref{psiI}, \eqref{psiII} and 
\eqref{psiIII}. With $\xi=\pm$, we find the relations
\begin{equation}\label{symrel1}
C^{(n,k)}_I = \xi\left( C^{(n,-k)}_{III}\right)^*,\quad
C^{(n,k)}_{\pm,II} = \xi \left( C^{(n,- k)}_{\mp,II} \right)^*.
\end{equation}
Let us then choose the overall phase of each eigenstate such that
all $C_I^{(n,k)}$ are real-valued.  Hence Eq.~(\ref{symrel1}) implies that
all coefficients  in the vector 
${\bm C}_{n,k}=\left(C^{(n,k)}_I,C^{(n,k)}_{+,II},
C^{(n,k)}_{-,II},C^{(n,k)}_{III}\right)^T$ are also
real-valued, which in turn confirms that $\phi_{n,k}$ 
and $\chi_{n,k}$ in Eq.~\eqref{free-solution} can indeed be chosen real-valued.

Using the real-valued coefficient vector ${\bm C}_{n,k}$,
the matching conditions can be written in compact form as 
${\bm M}_{n,k}(E) \cdot {\bm C}_{n,k}= 0$,
where the $4\times 4$ matrix ${\bm M}_{n,k}(E)$ 
is easily read off from Eqs.~\eqref{psiI}, \eqref{psiII} and 
\eqref{psiIII}.  The eigenenergies $E=E_{n,k}$ then follow from the condition
\begin{equation}\label{detM}
{\rm det}\ {\bm M}_{n,k}(E)=0,
\end{equation}
and the normalized eigenstates $\psi_{n,k}(x)$ are determined 
from the corresponding
eigenvectors ${\bm C}_{n,k}$.  In general, the solutions to Eq.~\eqref{detM} 
have to be obtained by using numerical root-finding methods \cite{tarun}.

\section{On the limit $d\to \infty$ }\label{appb}

Here we briefly discuss the large-$d$ behavior of the
single-particle solutions in Sec.~\ref{sec2a}. 
In effect, increasing the MGW width $d$ is a way to 
reverse the direction of the magnetic field, 
$B\hat {\bm e}_z\to -B\hat {\bm e}_z$,
in the bulk of the sample.
Eventually, as $d\to \infty$, all eigenstates must approach to Landau levels 
with index $\tilde n\in\mathbb{Z}$ of the time-reversed system, 
\begin{equation}\label{zeroLL}
\psi_{\tilde n,k}(x)= \frac{1}{\sqrt{1+{\rm sgn}(|\tilde n|)}}
\left(\begin{array}{c} \varphi_{|\tilde n|}(x+k)\\
i\, {\rm sgn}(\tilde n) \varphi_{|\tilde n-1|}(x+k)\end{array}\right),
\end{equation}
where the $\varphi_n$ are normalized eigenstates of the 1D harmonic oscillator
and ${\rm sgn}(n)=(1,0,-1)$ for $(n>0,n=0,n<0)$.
In particular, a zeroth Landau level must arise, $\tilde n=0$, where
now only the upper spinor component is nonzero as 
compared to Eq.~\eqref{zeropsi}.
This development is nicely tracked from the inset of Fig.~\ref{fig3} together 
with Fig.~\ref{fig2}. Indeed, as $d$ increases, the $n=\pm 1$ snake levels 
successively flatten in order to ultimately join at zero energy. 
(Note that the avoided crossing in Fig.~\ref{fig2} shifts 
towards bigger $|k|$ values with increasing $d$.)
Eventually, the former $n=\pm 1$ snake levels merge at zero to form
the new zeroth Landau level, see Eq.~\eqref{zeroLL} with 
$\tilde n=0$.  In fact, using properties of the parabolic cylinder functions
\cite{gradst,abramowitz}, one finds [see Eq.~\eqref{psiII}]
\begin{equation}
\psi_{n,k}(x) \to\pi^{-1/4}e^{-(x-k)^2/2} \left( \begin{array}{c}
1\\0\end{array}\right),  \quad E\to 0.
\end{equation}

\section{Form factor symmetries}\label{appc}

In this Appendix, we study general properties of
 the form factors defined in Eq.~\eqref{bdef}.
Using Eq.~\eqref{symrel2}, we find that they obey the symmetry relations 
\begin{eqnarray}\label{symrelF1}
{\cal F}^{(0,n)}_{k,q}(k_x) &=& - {\cal F}^{(0,-n)}_{k,q}(k_x) ,\\
{\cal F}^{(n,n')}_{k,q}(k_x) &=& {\cal F}^{(-n,-n')}_{k,q}(k_x),\nonumber\\ 
{\cal F}^{(n,n')}_{k,q}(k_x) &=& {\cal F}_{k+q,-q}^{(n',n)}(k_x),\nonumber\\
{\cal F}^{(n,n')}_{k,q}(k_x) &=& 
(-1)^{n+n'}{\cal F}_{-k,-q}^{(n,n')}(-k_x) ,\nonumber 
\end{eqnarray}
where the first two relations only hold for $n\ne 0$ and $n'\ne 0$, while
the last two are valid for arbitrary $n,n'$. Furthermore, as a result of  
\begin{equation}
{\cal F}_{k,q}^{(n,n')}(k_x)= \left( 
{\cal F}_{k,q}^{(n,n')}(-k_x) \right)^*,
\end{equation}
all Coulomb matrix elements in Eq.~\eqref{coul} are real-valued.
Using Eq.~\eqref{symrelF1}, we find that they obey the symmetry relations
quoted in Eq.~\eqref{symrel}.


\begin{thebibliography}{99}
\bibitem{footcurr}
We note that along the $x$-direction, with $\sigma_y\to \sigma_x$ in 
Eq.~\eqref{currmatrel}, perpendicular to the MGW,
the diagonal current matrix elements vanish identically since 
$\phi_{n,k}$ and $\chi_{n,k}$ are real-valued.
\bibitem{foot}
This rule is obtained from 
$\lim_{\omega\to 0} dZ_{s=\pm}(\omega)/d\omega=0$
with $Z_s(\omega)= \int dE (\ldots)
\tilde {\cal G}^{(-s,s)}_{n,k} (E+s\omega) I_{n,k;n',k'} \tilde {\cal G}_{n',k'}^{(s,-s)}(E)$,
which in turn follows from Eqs.~\eqref{zerocur} and \eqref{gkf}.
\end{thebibliography}
\end{document}